\documentclass[prl,showpacs,twocolumn,amsmath,nofootinbib,amssymb]{revtex4}

\usepackage{bm}
\usepackage{mathrsfs}
\usepackage{amssymb}
\usepackage{amsmath}
\usepackage{amsfonts}
\usepackage{color}
\usepackage{ulem}
\usepackage{cancel}

\usepackage[]{graphicx}
\begin{document}
\newcommand{\half}{\frac{1}{2}}
\title{A no-go theorem for $\Psi-$anomic models under the restricted ontic indifference assumption}
\author{Aur\'elien Drezet$^{1}$}
\address{$^1$Univ. Grenoble Alpes, CNRS, Grenoble INP, Institut Neel, F-38000 Grenoble, France}

\email{aurelien.drezet@neel.cnrs.fr}
\begin{abstract}
 We address the question of whether a non-nomological (i.e., anomic) interpretation of the wavefunction is compatible with the quantum formalism. After clarifying the distinction between ontic, epistemic, nomic and anomic models we focus our attention on two famous no-go theorems due to Pusey, Barrett, and Rudolph (PBR) on the one side and Hardy on the other side which forbid the existence of anomic-epistemic models. Moreover, we demonstrate that the so called restricted ontic indifference introduced by Hardy induces new constraints.  We show that after modifications the Hardy theorem actually rules out all anomic models of the wavefunction assuming only restricted ontic indifference and preparation independence. 
\end{abstract}
\maketitle

\begin{figure}[h]
\begin{center}
\includegraphics[width=6cm]{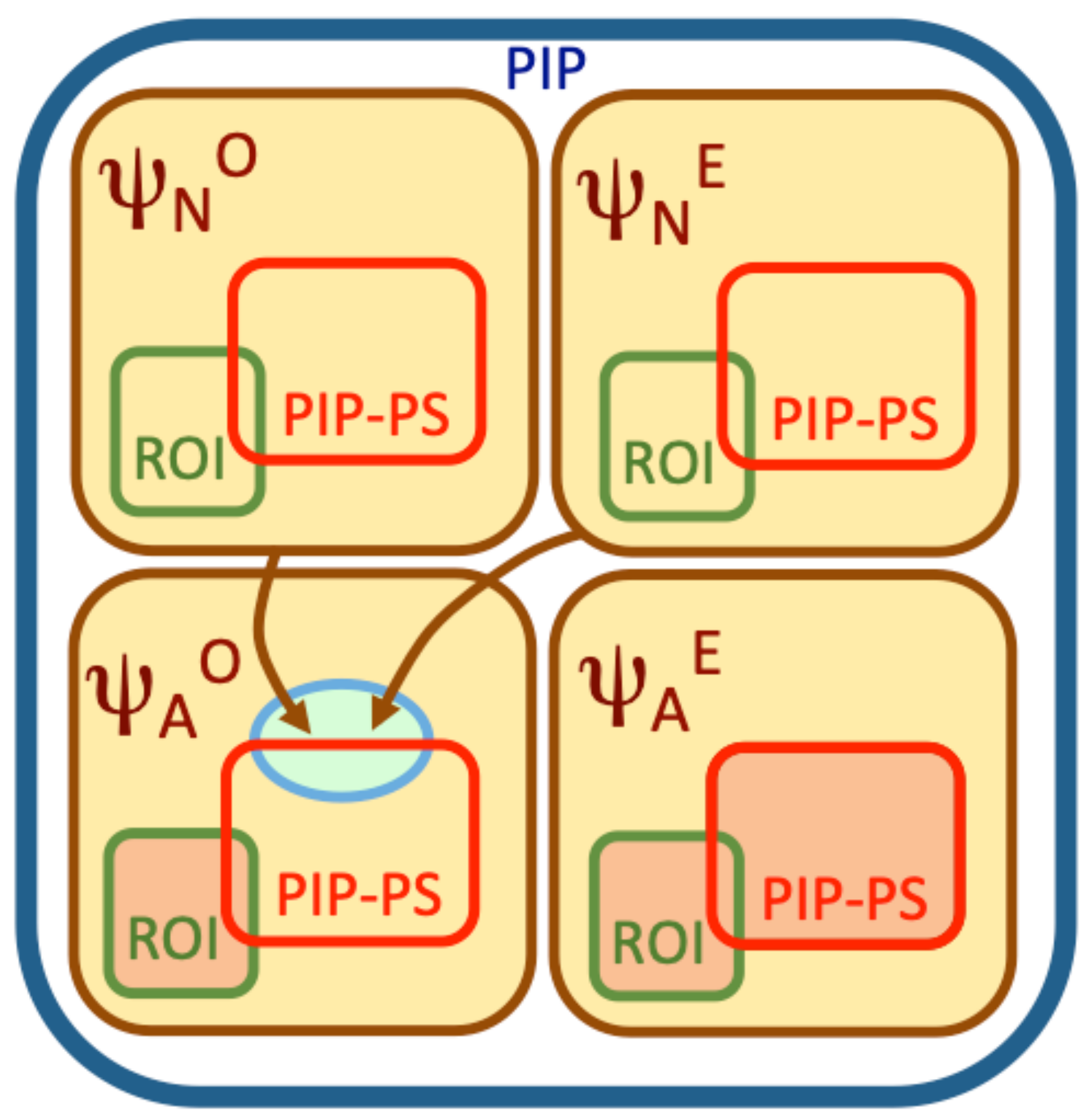} 
\caption{Classification of the different ontological models assuming the PIP (see text). $\psi_N^O$ and $\Psi_N^E$ models can be transformed (see arrows) into models of the $\psi_A^O$ class (blue ellipse domain). Models satisfying  PIP-PS (red boxes) and ROI (green boxes) are also represented. Dark orange regions are prohibited by no-go theorems: $\psi_A^E$ models assuming PIP-PS are excluded by the PBR theorem~\cite{Pusey}, whereas the ROI domain in the $\Psi_A$ regions are excluded by the H-ROI (II) theorem derived in this work. Note, that the models prohibited by the original H-ROI theorem~\cite{Hardy} are inside the ROI subdomain of the  $\psi_A^E$ class. } \label{figure0}
\end{center}
\end{figure}

\indent The nature of the wavefunction $\Psi$ is the topic of strong debates and controversies since its introduction in quantum physics in the 1920-30's~\cite{Valentini}. In the recent years the debate became more technical and focussed on the so called $\psi-$epistemic (here after $\psi^E$) vs $\psi-$ontic  (here after $\psi^O$) distinction (see Fig.~\ref{figure0}). This terminology originally introduced by Harrigan and Spekkens~\cite{Harrigan} has the following prerequisite: First, write $\mathbb{P}^M_\Psi(\alpha):=|\langle\alpha|\Psi\rangle|^2$  the quantum probability (Born's rule) for observing the outcome $\alpha$, i.e., associated with the state $|\alpha\rangle $ during a measurement $M$, when the quantum system belonging to the Hilbert space $\mathcal{H}$ is prepared in the $|\Psi\rangle$ state. Assuming the existence of underlying hidden-variables, or more generally `ontic states', $\lambda\in \Lambda$ we write with J.~S.~Bell~\cite{Bell}:
\begin{eqnarray}
\mathbb{P}^M_\Psi(\alpha)=\int_\Lambda  \mathbb{P}^M_\Psi(\alpha|\lambda)\mathbb{P}^M_\Psi(\lambda)d\lambda\label{eq1}
\end{eqnarray}  where  $\mathbb{P}_\Psi(\alpha|\lambda)$ is the response, or indicator, function for the hidden-variable theory considered, i.e., the probability to record the outcome $\alpha$ conditioned on the hidden variable  value $\lambda$. Similarly, $\mathbb{P}^M_\Psi(\lambda)$ denotes the (density of) probability for the hidden-variables \textit{to be} in the state characterized by $\lambda$. These probabilities are fulfilling the obvious normalization conditions:  $\sum_\alpha\mathbb{P}^M_\Psi(\alpha|\lambda)=1$ (where the sum is taken over the complete measurement basis), and  $\int_\Lambda  \mathbb{P}^M_\Psi(\lambda)d\lambda=|\langle\Psi|\Psi\rangle|^2=1$. Here we also assume the \textit{preparation independence postulate} (PIP):
\begin{eqnarray}
\mathbb{P}^M_\Psi(\lambda):=\mathbb{P}_\Psi(\lambda)\label{eq0}
\end{eqnarray}
\\
\indent According to~\cite{Harrigan}, the theory is $\psi^0$ iff for every pair of states $|\Psi_1\rangle,|\Psi_2\rangle \in\mathcal{H}$  we have 
\begin{eqnarray}
\mathbb{P}_{\Psi_1}(\lambda)\mathbb{P}_{\Psi_2}(\lambda)=0.\label{eq2}
\end{eqnarray} Otherwise the theory will be said to be $\psi^E$. This terminology stresses the classical intuition that for a  $\psi^0$ theory the hidden variables distributions associated with different quantum states $\Psi_i:=\Psi_1, \Psi_2,...$ must have disjoints supports $\Lambda_{\Psi_i}$ (i.e., no overlap) in the $\Lambda$-space, whereas an overlap should be generally allowed for a $\psi^E$ theory.\\
\indent  In classical physics the density of probability  $\mathbb{P}(q,p)$ in the phase space is an epistemic property and we can always find two distributions such that $\mathbb{P}_1(q,p)\mathbb{P}_2(q,p)\neq 0$. The question is thus to see if the same holds true in quantum mechanics, i.e., if $\Psi$ is just a label for the probability distributions (in that case we should have a $\psi^E$ theory) or if $\Psi$ has a more fundamental meaning as intuited from interference phenomena (in that case we should have a $\psi^O$ theory).\\ 
\indent  In this letter we consider two remarkable such attempts for clarifying this $\psi^O, \psi^E$ ambiguity:  The Pussey-Barrett-Rudoph (PBR)  theorem~\cite{Pusey,Pusey2}, and the Hardy `restricted ontic-indifference' (H-ROI) theorem~\cite{Hardy} against $\psi^E$ models~\cite{noteb}. The goal here is not to review these important results but to clarify their impacts and limitations. Moreover, we show that the H-ROI theorem is not just a variant of the PBR theorem but actually can be modified into a stronger no-go result against the existence of many `natural' ontological models.\\
\indent The PBR theorem (see \cite{Leifer,Boge} for reviews) shows that assuming an additional preparation independence postulate for product states (PIP-PS) ontological models must be $\psi^O$, i.e., $\psi^E$ theories conflict with quantum mechanics.  For non orthogonal states~\cite{notea} $\langle\Psi_1|\Psi_2\rangle\neq 0$ the result is derived by assuming product states like $|\Psi_n\rangle^{(A)}\otimes|\Psi_m\rangle^{(B)}\in \mathcal{H}^{(A)}\otimes \mathcal{H}^{(B)}$ (where $n,m=1$ or 2 and $A,B$ label two copies of the same Hilbert space). The PIP-PS reads  $\mathbb{P}_{\Psi_n\otimes\Psi_m}(\lambda^{(A)},\lambda^{(B)})=\mathbb{P}_{\Psi_n}(\lambda^{(A)})\mathbb{P}_{\Psi_m}(\lambda^{(B)})$ (here we introduced a Cartesian product hidden-variables space $\Lambda^{(A)}\times\Lambda^{(B)}$). In \cite{Pusey} measurement protocols involving antidistinguishable product states~\cite{Leifer} are proposed in order to justify Eq.~\ref{eq2} for every pairs of states in $\mathcal{H}$.\\
\indent Several comments  must be done concerning the PBR theorem. First, observe that the derivation also assumes 
\begin{eqnarray}
\mathbb{P}^M_\Psi(\alpha|\lambda)=\mathbb{P}^M(\alpha|\lambda).\label{eq3}
\end{eqnarray} This rather innocuous axiom Eq.~\ref{eq3} was implicit in \cite{Pusey} (this was already pointed out in \cite{Drezet1,Drezet2,Drezet3,Drezet4} and independently in~\cite{Hall,Fineart}). Moreover, it plays a fundamental role since it implies  $\psi-$\textit{independence at the law level}, or in other words that  $\Psi$ is not involved in the dynamics of $\lambda$: It is a $\psi-$anomic theory  (here after $\psi_A$). We stress that the well-known de Broglie-Bohm (dBB) hidden-variables theory~\cite{Holland,Bohm}, which is empirically equivalent to standard quantum mechanics, violates conditions given by Eq.~\ref{eq3}, i.e., $\psi_A$~\cite{Drezet4}. Therefore, such a theory is said to be `nomological'~\cite{Bohm} or  $\psi-$nomic (here after $\psi_N$).\\
\indent Now, what the PBR  no-go theorem really says is that:\\
\indent\textbf{PBR theorem}-- Assuming PIP-PS there is no $\psi_A^E$ theory.\\
\indent In other words: A $\psi_A$ theory can not be $\psi^E$ and must therefore be $\psi^O$ (see Fig.~\ref{figure0}).  The theorem says nothing about $\psi_N$ models (i.e., about the existence of $\psi_N^O$ and $\psi_N^E$ models) and therefore the derivation \cite{Pusey} can not run for these cases (e.g., the dBB theory is $\psi_N^E$~\cite{supplement}).  We believe that the unfortunate choice of not clearly distinguishing  between dynamics and statistics in the terminology of Harrigan and Spekkens was responsible for many confusions surrounding the PBR theorem~\cite{crit}. Furthermore, the distinction between $\psi^O$  and $\psi_N$ removes terminological ambiguities~\cite{notespe} and clarify the role of Eq.~\ref{eq3}.\\
\indent A second comment concerns the fact that it is always possible to extend the hidden-variables space $\Lambda$ to include a supplementary variable $\tau_\Psi\in \Gamma$ isomorphic to the wavefunction $|\Psi\rangle$: For example in SU(2) a spinor on the unit sphere is characterized by angles $\vartheta,\varphi$ on the Bloch sphere. A more general methods is given in \cite{Belma,Drezet3,foot1}. The new ontic space $\Lambda\times \Gamma$ allowed Harrigan and Spekkens to distinguish between $\Psi-$complete models where only $\Gamma$ is considered and  $\Psi-$supplemented models involving $\Lambda\times \Gamma$. Moreover, Eq.~\ref{eq1} can be rewritten
\begin{eqnarray}
\mathbb{P}^M_\Psi(\alpha)=\int_{\Lambda\times \Gamma}  \mathbb{P}^M(\alpha|\mu)\mathbb{\tilde{P}}_\Psi(\mu)d\mu\label{eq5}
\end{eqnarray}   where by definition $\mu:=(\lambda,\tau_{\Phi})$, and $\mathbb{P}(\alpha|\mu):=\mathbb{P}_{\Phi}(\alpha|\lambda)$ (with $\tau_{\Phi}\leftrightarrow|\Phi\rangle$ and $|\Phi\rangle\in \mathcal{H}$). The density of probability $\mathbb{\tilde{P}}_\Psi(\mu)$ is by definition~\cite{foot2}
\begin{eqnarray}
\mathbb{\tilde{P}}_\Psi(\mu):=\mathbb{P}_\Psi(\lambda)\delta(\tau_\Phi-\tau_\Psi).\label{eq6}
\end{eqnarray}  
\indent Moreover, from Eqs.~\ref{eq5} and \ref{eq6} we now have a $\psi_A^O$ theory. Such a model trivially satisfies the PBR theorem since $\mathbb{\tilde{P}}_{\Psi_1}(\mu)\mathbb{\tilde{P}}_{\Psi_2}(\mu)=0$ $\forall \mu \in\Lambda\times \Gamma$ and for every pairs $|\Psi_1\rangle,|\Psi_2\rangle \in\mathcal{H}$.  Therefore, by adding an hidden variable $\tau_{\Phi}$ to $\lambda$ we can always transform any $\psi_N^O$ or $\psi_N^E$ model into a $\psi_A^O$ theory (see Fig.~\ref{figure0}).  We emphasize that even if this new models are mathematically and empirically equivalent to their parents they are however not ontologically equivalent since the new ontic space is now $\Lambda\times \Gamma$. \\ 
\indent This shed some new lights on a old debate surrounding the dBB theory~\cite{Bohm,Sole,Feintzeig}: Should the wavefunction be part of the ontology or should it better be considered as a nomological feature guiding the particles?  
We now see that these two approaches are mathematically and empirically equivalent, i.e., both agreeing precisely with the statistical predictions of quantum mechanics. The primitive ontology of particles in the $\Lambda$ space associated with a $\psi_N^E$ dBB ontology can be transformed into a new  $\psi_A^O$ theory in the $\Lambda\times \Gamma$ space where the wavefunction has now also an ontological nature. Therefore, at the end we have two different ontologies.\\  
\indent Having recapped this, it is clear that the PBR theorem must be supplemented by others assumptions in order to lead to physical conclusions on $\psi_A$ and $\psi_N$ theories. We believe that the goal can be partially reached using a modification of the original H-ROI theorem.\\
\indent The motivations for the H-ROI theorem~\cite{Hardy} is tied to the $\psi_N^E$ dBB particle ontology. Indeed, in the dBB theory the wavefunction in the configuration space transfers information from the environment to the particles and this in turn explains phenomena such as interferences and quantum-correlations. In the case of single-particle Mach-Zehnder interferences the particle after the first beam-splitter $BS_0$  follows necessarily one path (e.g., in arm $|0\rangle$). However, an `empty wave' \cite{Holland,Hardy1} must be included in the second arm (with state $|1 \rangle$) in order to give phase-information to the particle which in turn  determines its subsequent motion when crossing the second beam-splitter $BS_1$. Obviously, it seems very difficult to obtain this result without invoking a $\psi_N$ theory. The motivation of H-ROI is thus to justify this physical intuition by considering  a $\psi_A$ `particle-like' model. In such a $\psi_A$ ontology for localized hidden-variables we must invoke a form of locality:  \textit{Restricted ontic indifference} (ROI)~\cite{Hardy} expressing that any quantum operation made on the state $|0 \rangle$ and leaving it unchanged, doesn't impact the underlying hidden-variables $\lambda\in \Lambda_{0}$ in the ontic support of $|0 \rangle$ (otherwise the model would be $\psi_N$). We stress that Hardy \cite{Hardy}  also defined ontic indifference for all states but in the present work  we will limit our analysis to ROI leading to a restricted no-go result.\\
\begin{figure}[h]
\begin{center}
\includegraphics[width=8cm]{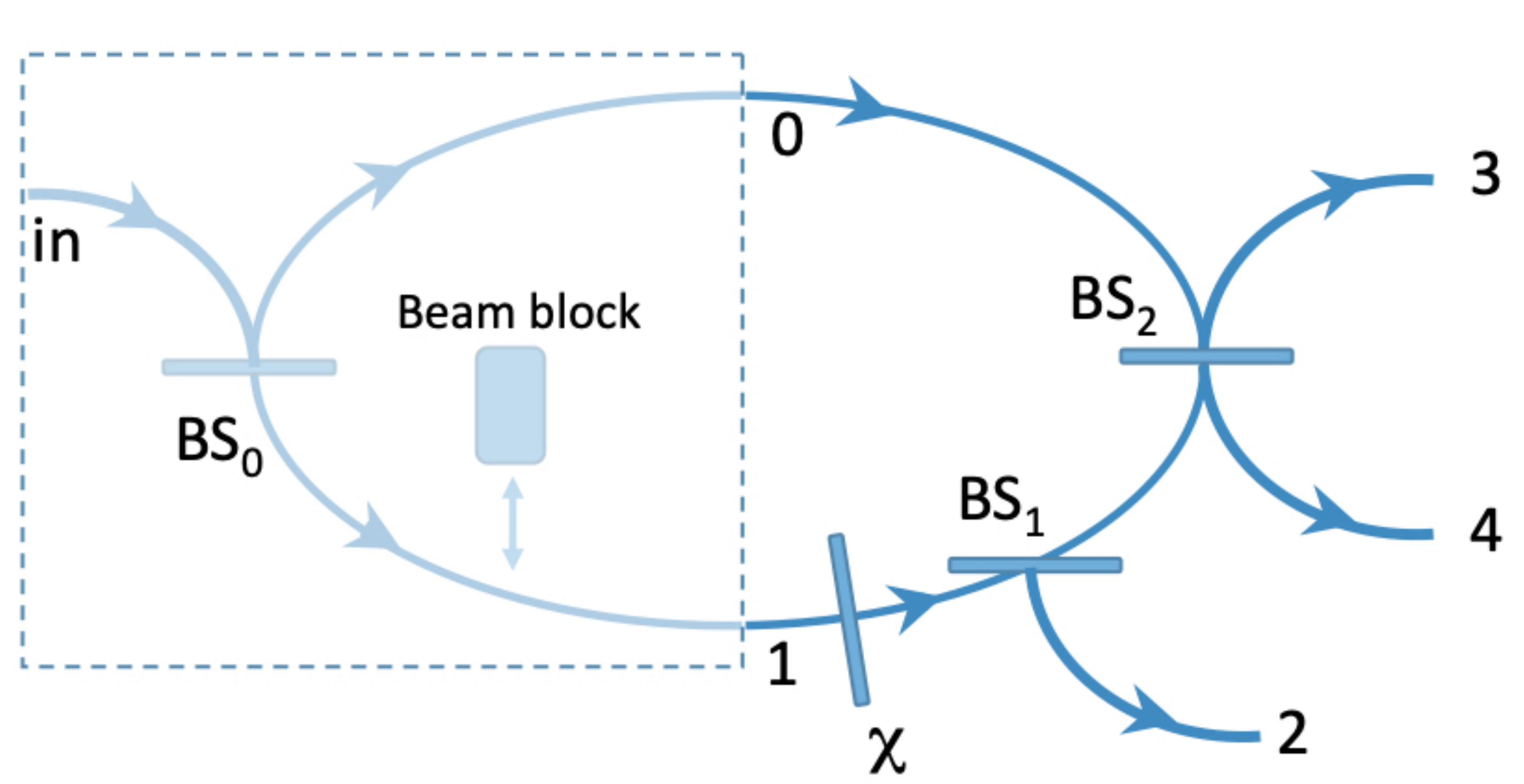} 
\caption{Principle of the interferometric proof for the H-ROI theorem~\cite{Hardy}. $BS_i$ ($i=0,1,2$) denote beam-splitters, and paths/gates are labeled as indicated in the figure (see main text). A phase-shifter ($\chi$) is introduced in path 1. The sketch also includes (light blue color region in the dashed box) the preparation stage (with a movable beam-blocker) which is used to modify the original proof given in \cite{Hardy}.} \label{figure1}
\end{center}
\end{figure}
\indent We here consider the `half' Mach-Zehnder sketched in  dark blue in Fig.~1. After a first beam-splitter BS$_0$ (shown in light blue in the dashed box of Fig.~1) a single electron beam has been prepared in the superposition  \begin{eqnarray}
|\Psi_+\rangle=a |0\rangle +b|1\rangle 
\end{eqnarray}  where $|0\rangle,|1\rangle$ are the two modes located in different arms, and   $a,b\in\mathbb{R}^+$ are normalized real positive amplitude coefficients ($a^2+b^2=1$, $b\geq a$, the phases are absorbed into the definition of $|0\rangle,|1\rangle$). In beam $1$ we add a wave-plate inducing a phase delay $\chi$:    $|1\rangle\rightarrow e^{i\chi}|1\rangle$. We thus obtain: 
\begin{eqnarray}
|\Psi_+\rangle\underset{\chi}{\rightarrow}a |0\rangle +e^{i\chi}b|1\rangle.
\end{eqnarray} In particular, if $\chi=\pi$ we get the state 
\begin{eqnarray}
|\Psi_+\rangle\underset{\chi=\pi}{\rightarrow} |\Psi_-\rangle=a |0\rangle -b|1\rangle 
\end{eqnarray} with $\langle\Psi_+|\Psi_-\rangle=a^2-b^2$. A beam-splitter  (BS$_1$) is subsequently added in beam 1 (see Fig.~1) and we have the transformation
\begin{eqnarray}
a|0\rangle +e^{i\chi}b|1\rangle\underset{\textrm{BS}_1}{\rightarrow}a |0\rangle +e^{i\chi}T b|1\rangle-e^{i\chi}Rb|2\rangle
\end{eqnarray}  where $T$ and $R=\sqrt{1-T^2}$ are the transmission and reflectivity amplitudes respectively (the minus sign comes from unitarity). 
In the following we impose $T=a/b$ and we finally introduce a last 50/50 beam splitter (BS$_2$)  with input modes $|0\rangle,|1\rangle$ and outcomes $|3\rangle,|4\rangle$. This leads to the transformation:
\begin{eqnarray}
a |0\rangle +e^{i\chi}Tb|1\rangle-e^{i\chi}Rb|2\rangle\nonumber\\
=a |0\rangle +e^{i\chi}a|1\rangle-e^{i\chi}\sqrt{b^2-a^2}|2\rangle\nonumber\\
\underset{\textrm{BS}_2}{\rightarrow}\frac{a}{\sqrt{2}}(1+e^{i\chi}) |3\rangle +\frac{a}{\sqrt{2}}(1-e^{i\chi})|4\rangle\nonumber\\-e^{i\chi}\sqrt{b^2-a^2}|2\rangle.\label{eq11}
\end{eqnarray}
We consider two particular cases: If $\chi=0$ we have 
\begin{eqnarray}
|\Psi_+\rangle\underset{\chi=0}{\rightarrow} |\Psi_+\rangle\underset{\textrm{BS}_1,\textrm{BS}_2}{\longrightarrow} \sqrt{2}a|3\rangle -\sqrt{b^2-a^2}|2\rangle,
\end{eqnarray} and if $\chi=\pi$ we instead obtain
\begin{eqnarray}
|\Psi_+\rangle\underset{\chi=\pi}{\rightarrow} |\Psi_-\rangle\underset{\textrm{BS}_1,\textrm{BS}_2}{\longrightarrow} \sqrt{2}a|4\rangle +\sqrt{b^2-a^2}|2\rangle.
\end{eqnarray}
Suppose we have only a state in the $|0\rangle $ mode (for example by blocking the $|1\rangle $ gate just after BS$_0$). Letting the wave-plate (i.e., whatever $\chi$ is) and BS$_1$ in place in the empty path $|1\rangle $ doesnt affect beam $|0\rangle $ evolution which is only impacted by  BS$_2$.  We thus deduce   
\begin{eqnarray}
|\Psi_0\rangle:=|0\rangle\underset{\chi,\textrm{BS}_1,\textrm{BS}_2}{\longrightarrow} \frac{1}{\sqrt{2}}|3\rangle+\frac{1}{\sqrt{2}}|4\rangle.
\end{eqnarray}
Now, assuming a $\psi_A$ model satisfying the PIP and ROI we consider the hidden variable $\lambda\in\Lambda_{\Psi_+}$ in the ontic support of $|\Psi_+\rangle$. Since this is a $\psi_A$ model Eq.~\ref{eq3} holds true and we can define $\mathbb{P}^\chi(3|\lambda),\mathbb{P}^\chi(4|\lambda)$ where the superscript $\chi$ reminds that the experimental protocol, i.e., the response function, generally depends on the value $\chi$. Moreover, if $\chi=0$ we get 
\begin{eqnarray}
\mathbb{P}^{\chi=0}(4|\lambda)=0 &\forall\lambda\in\Lambda_{\Psi_+}\label{eq15}
\end{eqnarray}
and if $\chi=\pi$ we get 
\begin{eqnarray}
\mathbb{P}^{\chi=\pi}(3|\lambda)=0 &\forall\lambda\in\Lambda_{\Psi_+}.\label{eq16}
\end{eqnarray}Furthermore, for the particle prepared in state $|\Psi_0\rangle$ we have 
\begin{eqnarray}
\mathbb{P}(3|\lambda)+\mathbb{P}(4|\lambda)=1 &\forall\lambda\in\Lambda_{\Psi_0}\label{eq17}
\end{eqnarray} where the superscript $\chi$ of $\mathbb{P}_{\chi}(\alpha|\lambda)$ ($\alpha=3$ or 4) has been removed to agree with ROI. Finally, assuming with Hardy that $\lambda\in \Lambda_{\Psi_+}\cap \Lambda_{\Psi_0}$, we get from Eqs.~\ref{eq15},\ref{eq16}. 
\begin{eqnarray}
\mathbb{P}(3|\lambda)=0, \mathbb{P}(4|\lambda)=0 &\forall\lambda\in\Lambda_{\Psi_+}\cap \Lambda_{\Psi_0}.\label{eq18}
\end{eqnarray}  Eq.~\ref{eq18}, which is independent of $\chi$, conflicts with Eq.~\ref{eq17} and therefore we conclude~\cite{Hardy} that  $\Lambda_{\Psi_+}\cap \Lambda_{\Psi_0}=\emptyset$. In others words we get the result:\\
 \indent \textbf{H-ROI theorem}--$\psi_A$ models satisfying PIP and ROI can not be fully $\psi^E$.\\ \indent  As already mentioned this result is restricted to very particular states for which $b\geq a$, i.e., $|\langle\Psi_0|\Psi_+\rangle|^2\leq \frac{1}{2}$.  Its generalization to any pair of states would require to go beyond ROI~\cite{Hardy}.\\
\indent However, there is more in the H-ROI theorem. Indeed, going back to the preparation of state $|\Psi_0\rangle$ (i.e. in the light blue zone of Fig.~1) observe that we actually omitted one crucial step, that is, the transformation from the initial state $|\Psi_{in}\rangle$ existing before the initial BS$_0$ into the state $|\Psi_0\rangle$.  As we explained, this can easily be done by a beam-blocker removing the $|1\rangle$ mode  or by using an additional beam-splitter (i.e., to preserve unitarity and avoid further discussions about the particle absorption). In either case, it leads to the  evolution  
\begin{eqnarray}
|\Psi_{in}\rangle\underset{\textrm{BS}_0}{\longrightarrow}|\Psi_+\rangle\underset{abs.}{\longrightarrow} a|\Psi_0\rangle +|\textrm{rest}\rangle\label{prepar}
\end{eqnarray} where $|\textrm{rest}\rangle$ ($\lVert\textrm{rest}\rangle|\rVert^2=b^2$) is the irrelevant part of the state absorbed or deviated by the device and $a|\Psi_0\rangle=a|0\rangle:=|\Psi'_0\rangle$ constitutes the prepared mode.  Here, comes the issue:  Going back to Eq.~\ref{eq17} we must now have  
\begin{eqnarray}
\mathbb{P}(3|\lambda)+\mathbb{P}(4|\lambda)=1 &\forall\lambda\in\Lambda_{\Psi_{in}}[\Psi_0]\label{eq18b}
\end{eqnarray} where $\Lambda_{\Psi_{in}}[\Psi_0]\subset\Lambda_{\Psi_{in}}$ is the subset of $\Lambda_{\Psi_{in}}$ leading to the preparation of mode $|\Psi'_0\rangle$.  Again, a form of ROI was used (see below for comments). Moreover, by comparing with Eqs.~\ref{eq15}, \ref{eq16} (but with $\Lambda_{\Psi_+}$ replaced by $\Lambda_{\Psi_{in}}$) we get a contradiction: $\mathbb{P}(3|\lambda)+\mathbb{P}(4|\lambda)$ must equal zero and one at the same time $\forall\lambda \in\Lambda_{\Psi_{in}}[\Psi_0]\subset\Lambda_{\Psi_{in}}$ (this is a sketch of the proof assuming determinism; a more complete derivation is given in \cite{supplement}). Therefore, we have no other alternative than to abandon the $\psi_A$ model (see Fig.~\ref{figure0}). This leads to the main result of this article:\\ 
\indent \textbf{H-ROI (no-go) theorem (II)}--PIP and ROI together conflict with $\psi_A$ theories.\\ 
\indent Several remarks must be done concerning this result: First, observe  that in \cite{Hardy} no preparation stage Eq.~\ref{prepar} was involved since the motivation was to justify the PBR conclusion from different hypotheses (i.e., PIP and ROI instead of PIP-PS).  On the contrary, for our deduction Eq.~\ref{prepar} is key. Without relying on the PIP-PS, we actually precise the PBR theorem by showing that if we assume a  $\psi_A$ model PIP, and ROI then we necessarily run into a contradiction: Neither  $\psi_A^E$ nor $\psi_A^O$ models are therefore allowed.\\
\indent We point out that the definition of ROI used here is weaker than in \cite{Hardy}. Indeed,  Eq.~\ref{eq17} shows that the key idea is to refute the existence of an empty wave~\cite{Holland,Hardy1} and therefore if we know which path the particle is going along  (i.e., $|0\rangle$ due to the presence of the beam-blocker) the empty path not taken and what is inside it (i.e., the wave-plate in path $|1\rangle$) have no influence on the indicator function $\mathbb{P}(4|\lambda)$ and $\mathbb{P}(3|\lambda)$. But note that in our theorem H-ROI(II), the hidden variables $\lambda$ are defined before the wave-packet $|\Psi_{in}\rangle$ interacts with the device. The condition Eq.~\ref{eq18b} is thus more dynamics than in \cite{Hardy} and exploits the $\psi_A^E$ nature of the ontological models considered.  We note \textit{en passant} that our analysis of the preparation procedure shows some interesting connections with the notion of state update recently discussed in \cite{Ruebeck}. It should be interesting to further investigate this connection. We also remind that  we didn't here considered the broader framework of ontic indifference for all quantum states discussed in \cite{Hardy} (and in \cite{Patra} in relation with a continuity assumption).  However, since we already ruled out ROI for $\psi_A$ models this casts some doubts on the physical pertinence of a broader framework. This, clearly, should be  the subject of further work.\\ 
\indent Furthermore, we stress that if we start with a $\psi_N$ model, e.g., like the dBB theory, and if we supplement the model with a $\tau_\Phi\in \Gamma$ vector we will not in general be able to satisfy ROI since  the wavefunction that is now part of the ontological space $\Lambda\times\Gamma$ is a highly delocalized hidden variable (e.g., in modes $|\Psi_\pm\rangle$).  The derivation presented here could not run.  This again shows the importance of distinguishing between different mathematically equivalent frameworks (like dBB theory being either $\psi_N^E$ or $\psi_A^O$) when we apply physical principles such as ROI. Moreover, this shows that the dBB theory can not be ruled out by our theorem  prohibiting $\psi_A$ models with ROI. Indeed, either the  dBB theory is $\psi_N^E$, and agrees with ROI, or it is $\psi_A^O$ (in the $\Lambda\times \Gamma$ space) and doesn't agree with ROI. In each case there is no conflict with our no-go theorem. \\
\indent Finally, remark that whereas ontic indifference is a natural hypothesis for spatial degrees of freedom it is not a mandatory hypothesis.  For example, ROI is violated in the $\psi_A^E$ toy model proposed by Spekkens \cite{Spekkens,Hardy,Leifer}. Furthermore, dBB models for bosonic quantum fields \cite{Holland} using a wavefunctional representation $\Psi([\phi(\mathbf{x})],t)$, where $\phi(\mathbf{x}):=\lambda$ is a continuous field playing the role of an hidden-variable, also  generally disagree with ROI.\\   
\indent To conclude, after introducing a general terminology involving $\psi_A$ and $\psi_N$ models together with the more traditional $\psi^E$ and $\psi^O$ models used in the original Harrigan/Spekkens framework, we emphasized the fact that the PBR theorem only prohibits the existence of $\psi_A^E$ hidden-variables theories. $\psi_N$ models in general, and $\psi_N^E$ models in particular, are not forbidden.  The H-ROI theorem was subsequently analyzed in this framework and a stronger theorem: H-ROI(II) was derived which is proving the incompatibility of PIP, ROI and  $\psi_A$ theories. Altogether, this hierarchy of theorems imposes strong constraints on future hidden-variables models and opens new exciting questions concerning $\psi_N$ and $\psi_A$ models. In particular, it lets open the possibilities: (i) To further develop $\psi_N$ models which, like the dBB theory, assumes ROI, or (ii) to modify drastically the usual space-time ontology by relinquishing ROI. This suggests some highly nonlocal wavefunctional $\psi_N$ and $\psi^0$ approaches but it could even save $\psi_A^E$ models by dropping the PIP~\cite{Pusey,Leifer,modela,modelb,more,mansfield,myrvold} or the free-choice assumption~\cite{Renner}.   
 
\section{appendix1} 
 
\indent In the $\psi_N$ dBB approach for point-like particles~\cite{Bohm} the hidden-variables are the positions of the $N$ particles with coordinates $\mathbf{x}_1,...,\mathbf{x}_N\in \mathbb{R}^3$ in the `real' 3D space. These are regrouped under a single super-vector $X:=[\mathbf{x}_1,...,\mathbf{x}_N]$ in the configuration space $\mathbb{R}^{3N}$ where the wavefunction $\psi(X,t)$  evolves. Furthermore, in the dBB theory the particles have a deterministic dynamics and we have very generally \begin{eqnarray}
\frac{d}{dt}X_\psi(t)=F_\Psi(X_\psi(t),t)\end{eqnarray} which characterizes a first-order dynamic belonging to the $\psi_N$ class.\\
\indent Moreover, the density of probability \begin{eqnarray}
\mathbb{P}_\Psi(X_\psi(t),t)=|\psi(X_\psi(t),t)|^2
\end{eqnarray} defines a hidden-variable probability density $\mathbb{P}_\Psi(\lambda)$ if we identify $\lambda$ with the vector $X_\psi(t_0)$ at an initial time $t_0$ (i.e., $\mathbb{P}_\Psi(\lambda):=|\psi(X_\psi(t_0),t_0)|^2$). But since two wavefunctions can overlap in the configuration space Eq.~\ref{eq3} is in general not valid  and the model is thus $\psi_N^E$.\\
\indent It is however remarkable  that both de Broglie and Bohm conceived  the wavefunction as an ontic field.  De Broglie wanted to elaborate a theory where the wave field was the primary enity   (the double solution theory) whereas Bohm considered   the wavefunction as a quantum potential $Q_\Psi$ acting upon the particles and fields.  In this empirically equivalent $\psi_A^O$ formulation it is Eq.~\ref{eq5} of the main article that must be used instead of Eq.~\ref{eq1}.     
\section{appendix2}

\indent The general proof of the contradiction starts with Eq. 1 and 3 of the main article:
\begin{eqnarray}
\mathbb{P}^M_{\Psi}(\alpha)=\textrm{Tr}[E^M_\alpha\rho_\Psi] =\int_\Lambda  \mathbb{P}^M_{\Psi}(\alpha|\lambda)\mathbb{P}_\Psi(\lambda)d\lambda\nonumber\\
=\int_\Lambda  \mathbb{P}^{M}(\alpha|\lambda)\mathbb{P}_\Psi(\lambda)d\lambda\label{eq1sup}
\end{eqnarray}  where beside the PIP we used the fact that for a $\psi_A$ model $\mathbb{P}^M_{\Psi}(\alpha|\lambda)=\mathbb{P}^{M}(\alpha|\lambda)$.  We added a label $M$ for the quantum measurement protocol considered. Furthermore according to quantum mechanics we also defined the quantum probability using a projector operator $E^M_\alpha$ for the measurement outcome $\alpha$ ($\sum_\alpha E^M_\alpha=1$) and the density matrix $\rho_\Psi=|\Psi\rangle\langle \Psi|$ at initial time (we use the Heisenberg representation).\\
\indent We assume a measurement sequence  $M_1,M_2$ on the system described by the wavefunction  $|\Psi\rangle$ and write $\alpha$, $\beta$ the outcomes of the first and second measurements respectively. After introducing the $\Lambda$ space the joint quantum probability associated with recording $\alpha$ and $\beta$ reads:
\begin{eqnarray}
\mathbb{P}^{M_2,M_1}_{\Psi}(\beta,\alpha)=\textrm{Tr}[E^{M_1}_\alpha E^{M_2}_\beta E^{M_1}_\alpha\rho_\Psi]\nonumber\\
=\int_\Lambda  \mathbb{P}^{M_2,M_1}(\beta,\alpha|\lambda)\mathbb{P}_\Psi(\lambda)d\lambda
\label{eq2sup}
\end{eqnarray} or equivalently
\begin{eqnarray}
\mathbb{P}^{M_2,M_1}_{\Psi}(\beta,\alpha)=\int_\Lambda  \mathbb{P}^{M_2,M_1}(\beta|\alpha,\lambda)d\mathbb{P}_\Psi(\alpha,\lambda)\nonumber\\
=\int_\Lambda  \mathbb{P}^{M_2,M_1}(\beta|\alpha,\lambda)\mathbb{P}^{M_1}(\alpha|\lambda)\mathbb{P}_\Psi(\lambda)d\lambda.
\label{eq3sup}
\end{eqnarray} 
After comparing Eq.~\ref{eq2sup} and Eq.~\ref{eq3sup} we obtain:
\begin{eqnarray}
\mathbb{P}^{M_2,M_1}(\beta|\alpha,\lambda)=\frac{\mathbb{P}^{M_2,M_1}(\beta,\alpha|\lambda)}{\mathbb{P}^{M_1}(\alpha|\lambda)}
\label{eq4sup}
\end{eqnarray} which obeys the usual normalization for the indicator function:\begin{eqnarray}
\sum_\beta\mathbb{P}^{M_2,M_1}(\beta|\alpha,\lambda)=\frac{\sum_\beta\mathbb{P}^{M_2,M_1}(\beta,\alpha|\lambda)}{\mathbb{P}^{M_1}(\alpha|\lambda)}=1.
\label{eq4sup}
\end{eqnarray} 
 \indent  For the present purpose we consider the deduction divided in 3 logical steps:\\ 
 \indent \textit{--Step (i)} As a first step (see Fig.~\ref{figure1} of the main article) the  detection of a particle at gates 3 or 4 after passing through arm $|0\rangle$ with the 
 beam-blocker in place and removing the wave propagating in arm $|1\rangle$. This corresponds to the sequence:
\begin{eqnarray}
|\Psi_{in}\rangle\underset{\textrm{BS}_0}{\longrightarrow}|\Psi_+\rangle\underset{abs.}{\longrightarrow} a|\Psi_0\rangle +|\textrm{rest}\rangle\nonumber\\
\underset{\chi,\textrm{BS}_1,\textrm{BS}_2}{\longrightarrow} \frac{a}{\sqrt{2}}(|3\rangle+|4\rangle)+|\textrm{rest}\rangle.
\label{prepar2}
\end{eqnarray}
As explained in the text of the main article the nature of state $|\textrm{rest}\rangle$ is not very crucial. It is here enough to have a clear which-path information  either using a beam-blocker or an entangling device.\\  
\indent We call $M_1$ the experiment: `The particle goes through $BS_0$ and is interacting with by the beam-blocker'. If the outcome $\alpha=$No the particle is stopped by the beam-blocker. If the outcome $\alpha=$Yes this corresponds to the preparation of state $|\Psi_0\rangle$. \\
\indent We call $M_2[\chi]$ the second part of the sequential experiment: `The particle goes through the interferometer with the wave-plate and $\textrm{BS}_1,\textrm{BS}_2 $ in place'. The different outcomes $\beta$ correspond to the label of the exit ports $\beta=2$, 3, or 4 (for completeness we also need to add a gate $\beta=\emptyset$ if the particle is stopped by the beam-blocker).\\
\indent In this experiment the joint probabilities $\mathbb{P}^{M_2[\chi],M_1}_{\Psi}(\beta,\alpha)$ corresponding to Eq.~\ref{prepar2} are  
\begin{eqnarray}
\mathbb{P}^{M_2[\chi],M_1}_{\Psi}(4,\alpha=\textrm{Yes})=\frac{a^2}{2},\nonumber\\
\mathbb{P}^{M_2[\chi],M_1}_{\Psi}(3,\alpha=\textrm{Yes})=\frac{a^2}{2}\nonumber\\
\mathbb{P}^{M_2[\chi],M_1}_{\Psi}(2,\alpha=\textrm{Yes})=0\nonumber\\
\mathbb{P}^{M_2[\chi],M_1}_{\Psi}(\emptyset,\alpha=\textrm{Yes})=0.
\label{eq5sup}
\end{eqnarray} We have also $\mathbb{P}^{M_2[\chi],M_1}_{\Psi}(\beta,\alpha=\textrm{No})=0$  if $\beta=2,3,4$ and  $\mathbb{P}^{M_2[\chi],M_1}_{\Psi}(\beta=\emptyset,\alpha=\textrm{No})=b^2$ but these are not interesting us here. All these probabilities are, of course, independent of the phase-shift $\chi$  and will be written $\mathbb{P}^{M_2[\cancel{\chi}],M_1}_{\Psi}(\beta,\alpha)$ in the following\\
\indent   Now, from Eq.~\ref{eq5sup} and Eq.~\ref{eq4sup} we get for $\lambda\in\Lambda_{\Psi_{in}}$
\begin{eqnarray}
\mathbb{P}^{M_2[\chi],M_1}_{\Psi}(2,\alpha=\textrm{Yes}|\lambda)=0,\mathbb{P}^{M_2[\chi],M_1}_{\Psi}(\emptyset,\alpha=\textrm{Yes}|\lambda)=0 \nonumber\\ \label{eq18bsupernew}
\end{eqnarray}
and 
\begin{eqnarray}
\mathbb{P}^{M_2[\cancel{\chi}],M_1}_{\Psi}(4,\alpha=\textrm{Yes}|\lambda)+
\mathbb{P}^{M_2[\cancel{\chi}],M_1}_{\Psi}(3,\alpha=\textrm{Yes}|\lambda)\nonumber\\=\mathbb{P}^{M_1}_{\Psi}(\alpha=\textrm{Yes}|\lambda)\label{eq18bnew}
\end{eqnarray} which generalizes Eq.~\ref{eq18b} of the main article.  In particular, for a deterministic hidden-variables theory we have $\mathbb{P}^{M_1}_{\Psi}(\alpha=\textrm{Yes}|\lambda)=1$ or 0.  If $\lambda\in\Lambda_{\Psi_{in}}[\psi_0]$ we have $\mathbb{P}^{M_1}_{\Psi}(\alpha=\textrm{Yes}|\lambda)=1$ and  Eq.~\ref{eq18bnew} reduces to Eq.~\ref{eq18b}:
\begin{eqnarray}
\mathbb{P}^{M_2[\cancel{\chi}],M_1}_{\Psi}(4,\alpha=\textrm{Yes}|\lambda)+
\mathbb{P}^{M_2[\cancel{\chi}],M_1}_{\Psi}(3,\alpha=\textrm{Yes}|\lambda)=1\nonumber\\ \label{eq18bnewnew}
\end{eqnarray} $\forall \lambda\in\Lambda_{\Psi_{in}}[\psi_0]$. Note that ROI was not  yet used in the reasoning so that Eq.~\ref{eq18bnewnew} is not yet exactly Eq.~\ref{eq18b}: This will require an other step discussed as step (iii) below.  \\
\indent  \indent \textit{--Step (ii)} As a second step we consider a different experiment where the beam-blocker has been removed. Instead of $M_1$ we now obtain the experiment $M_0$: `The particle goes through $BS_0$' which corresponds to the preparation of the state  $|\Psi_+\rangle$. The whole sequence $M_0$ followed by $M_2[\chi]$ leads to the state 
\begin{eqnarray}
|\Psi_{in}\rangle\underset{\textrm{BS}_0}{\longrightarrow}|\Psi_+\rangle
\underset{\chi,\textrm{BS}_1,\textrm{BS}_2}{\longrightarrow} \frac{a}{\sqrt{2}}(1+e^{i\chi}) |3\rangle \nonumber\\+\frac{a}{\sqrt{2}}(1-e^{i\chi})|4\rangle-e^{i\chi}\sqrt{b^2-a^2}|2\rangle,
\label{prepar2bb}
\end{eqnarray} and therefore to the probabilities
  \begin{eqnarray}
\mathbb{P}^{M_2[\chi],M_0}_{\Psi}(4)=a^2(1+\cos{\chi}),\nonumber\\
\mathbb{P}^{M_2[\chi],M_0}_{\Psi}(3)=a^2(1-\cos{\chi})\nonumber\\
\mathbb{P}^{M_2[\chi],M_0}_{\Psi}(2)=b^2-a^2.
\label{eq5supsup}
\end{eqnarray}
\indent  From Eq.~\ref{eq5supsup} and Eq.~\ref{eq4sup} we get for $\lambda\in\Lambda_{\Psi_{in}}$
\begin{eqnarray}
\mathbb{P}^{M_2[\chi=0],M_0}_{\Psi}(4|\lambda)=0,
\mathbb{P}^{M_2[\chi=\pi],M_0}_{\Psi}(3|\lambda)=0,\label{1516}
\end{eqnarray} and we have also
\begin{eqnarray}
\mathbb{P}^{M_2[\chi=0],M_0}_{\Psi}(3|\lambda)+\mathbb{P}^{M_2[\chi=0],M_0}_{\Psi}(2|\lambda)=1,\nonumber\\
\mathbb{P}^{M_2[\chi=\pi],M_0}_{\Psi}(4|\lambda)+\mathbb{P}^{M_2[\chi=\pi],M_0}_{\Psi}(2|\lambda)=1.\label{1516b}
\end{eqnarray} Eq.~\ref{1516} has the same meaning as Eqs.~\ref{eq15} and \ref{eq16} of the main article.\\
\indent  \indent \textit{--Step (iii)}  We must now introduce our definition of ROI. Going back to step (i), we want that the operations made in path 1 are inoperative for the dynamics if we already know that the particle went through path 0. More precisely, returning to  Eq.~\ref{eq18bnew} we want that the probabilities $\mathbb{P}^{M_2[\chi],M_1}_{\Psi}(4,\alpha=\textrm{Yes}|\lambda)$ and $\mathbb{P}^{M_2[\chi],M_1}_{\Psi}(3,\alpha=\textrm{Yes}|\lambda)$ are independent of what occurs in path 1.   This must be the case from ROI  and therefore we write our condition as:
 \begin{eqnarray}
\mathbb{P}^{M_2[\cancel{\chi}],M_1}_{\Psi}(4,\alpha=\textrm{Yes}|\lambda)=\mathbb{P}^{M_2[\chi],M_0}_{\Psi}(4|\lambda)\label{eq19nn}\\
\mathbb{P}^{M_2[\cancel{\chi}],M_1}_{\Psi}(3,\alpha=\textrm{Yes}|\lambda)=\mathbb{P}^{M_2[\chi],M_1}_{\Psi}(3|\lambda)\label{eq19nnn}
\end{eqnarray} where the condition expresses the fact that the beam-blocker doesn't change the dynamics (stochastic or deterministic) once we know the system selected path 0.\\
\indent   Moreover, from ROI the value of $\chi$ must also have no implication. Therefore, if we select $\chi=0$ in Eq.~\ref{eq19nn} and $\chi=\pi$ in Eq.~\ref{eq19nnn} we obtain from Eq.~\ref{1516} the result: 
\begin{eqnarray}
\mathbb{P}^{M_2[\cancel{\chi}],M_1}_{\Psi}(4,\alpha=\textrm{Yes}|\lambda)+
\mathbb{P}^{M_2[\cancel{\chi}],M_1}_{\Psi}(3,\alpha=\textrm{Yes}|\lambda)=0.\nonumber\\
\label{eq20f}
\end{eqnarray} This condition obviously contradicts Eq.~\ref{eq18bnew} since it leads to $\mathbb{P}^{M_1}_{\Psi}(\alpha=\textrm{Yes}|\lambda)=0$ which can not always be true (otherwise we would have $\mathbb{P}^{M_1}_{\Psi}(\alpha=\textrm{Yes})=0$ and $a^2$ at the same time).In particular for a deterministic model   this can not b true  $\forall\lambda \in\Lambda_{\Psi_{in}}[\Psi_0]\subset\Lambda_{\Psi_{in}}$ as discussed in the main article.\\
\indent We stress that ROI also leads to $\mathbb{P}^{M_2[\cancel{\chi}],M_1}_{\Psi}(2,\alpha=\textrm{Yes}|\lambda)=\mathbb{P}^{M_2[\chi],M_0}_{\Psi}(2|\lambda)$. Together with Eqs.~\ref{1516b} and \ref{1516} it yields $\mathbb{P}^{M_2[\cancel{\chi}],M_1}_{\Psi}(2,\alpha=\textrm{Yes}|\lambda)=1$ which obviously contradicts Eq.~\ref{eq18bsupernew}. All these deductions demonstrate the no-go theorem H-OI (II) discussed in the main article. 

\begin{thebibliography}{}
\bibitem{notea} For orthogonal states, i.e., $\langle\Psi_1|\Psi_2\rangle=0$, it is possible to find a measurement procedure with two outcomes $0,1$ such that $\mathbb{P}_{\Psi_1}(0)=0$ and $\mathbb{P}_{\Psi_2}(1)=0$.  Using Eqs.~\ref{eq1}, \ref{eq3} gives directly Eq.~\ref{eq2}, i.e., $\Lambda_{\Psi_1}\cap\Lambda_{\Psi_2}=\emptyset$.
\bibitem{notespe} We emphasize that the distinction between nomic and ontic interpretations of the dBB theory was already suggested by Harrigan and Spekkens~\cite{Harrigan} and let as an open problem `that is desserving of further scrutiny'.
\bibitem{foot1}The general idea of~\cite{Drezet3} is to consider a complete representation $\psi_n=\langle n|\Psi\rangle$ where $n\in \Delta$ is a label. We thus write $\tau_\Psi:=\{\psi_n=\psi'_n+i\psi''_n: n \in \Delta\}$ where $\psi'_n$ and $\psi''_n$ are the real and imaginary parts of $\psi_n$. We stress that similar ideas were already discussed by Beltrametti and  Bugajski for the $\psi-$complete case~\cite{Belma}.
\bibitem{foot2}Using the notations of \cite{Drezet3} we write $\delta(\tau_\Phi-\tau_\Psi):=\prod_n\delta(\Phi'_n-\Psi'_n)\delta(\Phi''_n-\Psi''_n)$ and $d\mu:=d\lambda\prod_nd\Phi'_nd\Phi''_n$.
\end{thebibliography}
\end{document}